# Symmetry-controlled SrRuO₃/SrTiO₃/SrRuO₃ magnetic tunnel junctions: Spin polarization and its relevance to tunneling magnetoresistance

Kartik Samanta[*] and Evgeny Y. Tsymbal[**]

*Department of Physics and Astronomy & Nebraska Center for Materials and Nanoscience,*
*University of Nebraska, Lincoln, Nebraska 68588, USA*

Magnetic tunnel junctions (MTJs), that consist of two ferromagnetic electrodes separated by an insulating barrier layer, have non-trivial fundamental properties associated with spin-dependent tunneling. Especially interesting are fully crystalline MTJs where spin-dependent tunneling is controlled by the symmetry group of wave vector. In this work, using first-principles quantum-transport calculations, we explore spin-dependent tunneling in fully crystalline SrRuO₃/SrTiO₃/SrRuO₃ (001) MTJs and predict tunneling magnetoresistance (TMR) of nearly 3000%. We demonstrate that this giant TMR effect is driven by symmetry matching (mismatching) of the incoming and outcoming Bloch states in the SrRuO₃ (001) electrodes and evanescent states in the SrTiO₃ (001) barrier. We argue that under the conditions of symmetry-controlled transport, spin polarization, whatever definition is used, is not a relevant measure of spin-dependent tunneling. In the presence of diffuse scattering, however, e.g. due to localized states in the band gap of the tunnel barrier, symmetry matching is no longer valid and TMR in SrRuO₃/SrTiO₃/SrRuO₃ (001) MTJs is strongly reduced. Under these conditions, the spin polarization of the interface transmission function becomes a valid measure of TMR. These results provide an important insight into understanding and optimizing TMR in all-oxide MTJs.

## 1. Introduction

Spintronics is an active research field that encodes information in electronic devices using spin degrees of freedom [1]. A commonly used spintronic device is a magnetic tunnel junction (MTJ) which consists of two ferromagnetic metal electrodes separated by a non-magnetic tunnel barrier [2-6]. Tunneling magnetoresistance (TMR) is the key functional property of MTJs. TMR is characterized by a change in resistance of the device when the relative magnetization of the two ferromagnetic electrodes is changed from parallel to antiparallel [7]. This resistance change serves as an ON/OFF ratio in a spintronic device and can reach several hundred percent, providing sufficient accuracy to read-out the magnetization state. The substantial TMR effect in MTJs enables them to be used as building blocks of magnetic random-access memories (MRAMs) for data storage and processing [8].

The TMR effect in MTJs is widely understood in terms of a spin-polarized tunneling current which is controlled by the relative magnetization orientation of the two FM electrodes. This mechanism is empirically quantified by Julliere's formula [2], $TMR = \frac{2p_1 p_2}{1 - p_1 p_2}$, where $p_1$ and $p_2$ are the spin polarization of the two ferromagnetic electrodes in an MTJ. Based on this formula, the electrodes with a larger spin polarization favor a larger TMR, and TMR is expected to be a function the transport spin-polarization. Quantitatively, however, the spin polarization is not uniquely defined and can be referred either to the uneven number of up-spin and down-spin electrons at the Fermi energy or to the unbalanced (spin-polarized) currents carried by electrons with opposite spin orientations [9]. Even in the latter case, the transport spin polarization appears to be different as determined from spin-dependent tunneling [3] or ballistic transmission [10,11] experiments.

Furthermore, in crystalline MTJs where the transverse wave vector is conserved in the tunneling process, an accurate description of spin-dependent transport is expected to consider symmetries of the incoming and outcoming Bloch states in the electrodes and evanescent states in the barrier [12]. In particular, matching of the majority-spin $\Delta_1$ band in the Fe (001) electrode to the $\Delta_1$ evanescent state in the MgO (001) barrier layer is known to be responsible for a large positive spin polarization and giant values of TMR predicted for crystalline Fe/MgO/Fe (001) MTJs [13,14]. Also, symmetry arguments explain a large negative spin polarization of electrons tunneling from ferromagnetic bcc Co (001) through SrTiO₃ (001) tunneling barrier [15] consistent with the experimental observations [16,17]. It is now commonly accepted that the transport spin polarization of MTJs is controlled by the ferromagnet/barrier pair rather than the ferromagnet alone, which can be understood in terms of the interface transmission function [18,19].

While these concepts are now well understood, there are not many experiments, apart from Fe/MgO/Fe (001) MTJs, where full crystallinity of MTJs is achieved and where the notions of symmetry matching could be explicitly verified for spin-dependent tunneling. Among different materials that can be utilized in MTJs, complex oxide ferromagnets and insulators are relevant because they can be grown epitaxially forming a single-crystalline full-oxide MTJ. For example, using SrTiO₃ (STO) as an insulating tunnel barrier, a TMR of 1800% at T = 4°K was demonstrated in LSMO/STO/LSMO MTJs [20], where a nominally half-metallic La$_{2/3}$Sr$_{1/3}$MnO₃ (LSMO) was used as an oxide electrode ferromagnet.

Among other magnetic oxides, SrRuO₃ (SRO) is interesting because it represents an itinerant ferromagnet that has perovskite structure with well-defined stoichiometry. SRO has a bulk Curie temperature of 160 K [21] with magnetism driven by Ru 4d electrons [22]. SRO has been extensively investigated [23], but recently gained an increased attention due to the emergent magnetic phenomena, such as anomalous and topological Hall effects, Weyl fermions, and topological spin textures associated with it [24-28]. Also, it represents a practical material to study current-induced magnetization switching [29] and perpendicular magnetic anisotropy tailored by the substrate [30]. In combination with perovskite oxide insulators, SRO can be used to explore the fundamental physics of spin-dependent tunneling in fully crystalline MTJs. For example, based on first-principles calculations, it was predicted that using ferroelectric $BaTiO_3$ as a tunnel barrier and SRO as electrodes in an MTJ leads to coexistent tunneling magnetoresistance and electroresistance effects [31]. Experimentally, STO was employed as a tunnel barrier in several experiments. Earlier studies have demonstrated small negative TMR effects ~2% in SRO/STO/LSMO MTJs [32], indicating a negative spin polarization of SRO of about 10% consistent with the preceding experimental measurements based on the Meservey-Tedrow technique [3] that utilized tunneling from SRO through STO to a superconductor [33]. Very recently, fully crystalline all-oxide SRO/STO/SRO MTJs have been grown and demonstrated much larger TMR ratios up to 25% [34], indicating a much higher spin polarization (~34% according to Julliere's formula) of SRO compared to that measured previously [32, 33]. These results indicate that the physics of spin-dependent tunneling in SRO/STO/SRO MTJs is not fully understood and requires further elucidation.

In this work, using first-principles quantum-transport calculations, we explore spin-dependent tunneling in fully crystalline SRO/STO/SRO (001) MTJs and predict TMR of nearly 3000%. We demonstrate that this giant TMR effect is driven by symmetry matching (mismatching) of the incoming and outcoming Bloch states in the SRO (001) electrodes and evanescent states in the STO (001) barrier. We argue that under the conditions of symmetry-controlled transport, spin polarization, whatever definition is used, is not a relevant measure of spin-dependent tunneling. In the presence of diffuse scattering, however, e.g. due to localized states in the band gap of the tunnel barrier, symmetry matching is no longer valid and TMR in SRO/STO/SRO (001) MTJs is strongly reduced. Under these conditions, the spin polarization of the interface transmission function becomes a valid measure of TMR. These results provide an important insight into understanding and optimizing TMR in all-oxide MTJs.

## 2. Methodology

The electronic structure calculations are carried out based on density functional theory (DFT) using the plane-wave projected augmented wave (PAW) method [35] as implemented in Vienna ab initio Simulation Package (VASP) [36,37]. We use the Perdew-Burke-Ernzerhof (PBE) [38] exchange-correlation functional within the generalized gradient approximation (GGA). For the self-consistent calculations, a plane-wave basis set with a plane-wave cutoff of 500 eV and a k-point mesh of 8×8×8 is used for the bulk cubic SRO and STO. Experimentally measured lattice constants of $a$ = 3.952 Å and 3.905 Å are assumed in the calculations for cubic bulk SRO [39] and STO [40], respectively. We consider an SRO/STO/SRO (001) heterostructure, consisting of 3 unit cells of SRO (001) on each side separated by 4 unit cells of STO (001). The in-plane lattice constant of the heterostructure is fixed to the lattice parameter of the cubic STO. A k-point mesh of 8×8×1 is used for self-consistent electronic structure calculations. The structural optimization is carried out using VASP maintaining the symmetry of the heterostructure. The positions of the atoms are relaxed toward equilibrium until the Hellman–Feynman forces become less than 0.01 eV/Å. For the Fermi surface calculations, we use the Wannier interpolation technique as implemented in the Wannier90 package [41]. A very dense k-point mesh of 100×100×100 is used to calculate the Fermi surface. The XCrySDen software is used to visualize the Fermi surface [42].

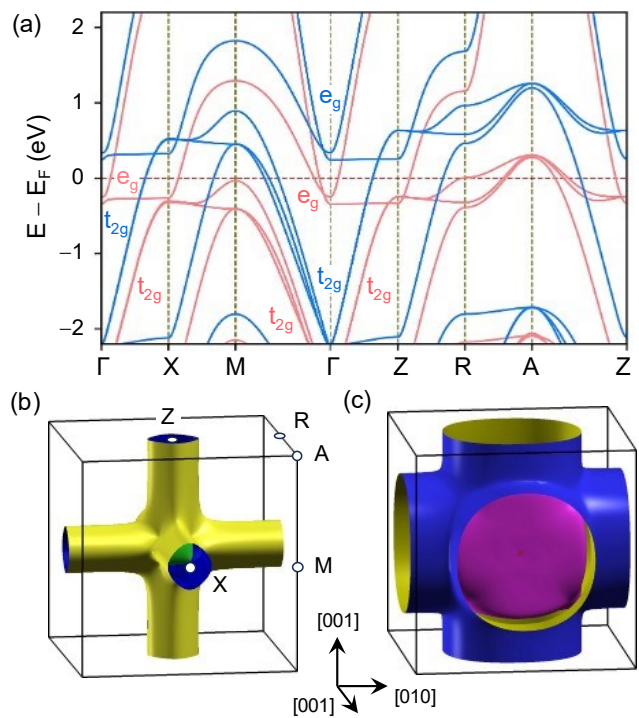

**FIG. 1** (a) Electronic band structure of bulk $SrRuO_3$ plotted along the high symmetry lines in the Brillouin zone for majority- (red lines) and minority- (blue lines) spin electrons. Dominant orbital contributions are indicated. (b,c) Fermi surfaces of the majority (b) and minority (c) spin of bulk $SrRuO_3$. Colors are used to aid the eye in delineating different sheets and different sides of the same sheet of the Fermi surface. High-symmetry points in the Brillouin zone are indicated.

Calculations of the transport properties are performed using the nonequilibrium Green's function formalism (DFT+NEGF approach) [43], as implemented in QuantumATK, Synopsys QuantumATK [44], using the atomic structures relaxed by VASP. In QuantumATK, the nonrelativistic Fritz-Haber-Institute (FHI) pseudopotentials are employed with a single-zeta-polarized basis, and a cut-off energy is set to 130 Ry. K-point meshes of 13×13×13 are used for bulk SRO and STO and 13×13×151 for SRO/STO/SRO MTJs. Transmission functions are calculated using k-point meshes of 401×401 in the two-dimensional Brillouin zone (2DBZ) of SRO and SRO/STO/SRO based MTJs.

Figures are plotted using VESTA [45], XcrySDen [42], gnuplot [46] and Python [47].

## 3. Spin-dependent properties of bulk SrRuO₃

Figure 1(a) shows the electronic band structure of bulk SRO, indicating itinerant ferromagnetic ground state with the exchange splitting of majority- and minority-spin bands of about 0.7eV. Owing to the strong delocalization of the Ru-4d electronic wave function, a large crystal field splitting at Ru-4d site stabilizes the low spin state. The calculated magnetic moment of 1.14 $\mu_B$ at the Ru sites is consistent with low spin state of the $Ru^{4+}$ in agreement with the earlier studies [48,49].

Figures 1(b) and 1(c) show the majority- and minority-spin Fermi surfaces of bulk SRO, each of them having several sheets [indicated by color in Figs. 1(b,c)]. An orbital analysis of the electronic states at the Fermi surface reveals that majority-spin states are mainly composed of the Ru $e_g$ orbitals. They are split into the $d_{x^2-y^2}$ states forming a band that represents a cross pattern of three corrugated tubes (the yellow surface in Fig. 1(b)) and the $d_{z^2}$ states forming a nearly spherical Fermi surface sheet of small radius [the green surface in Fig. 1(b) inside the tube]. On the contrary, the minority-spin states are mainly composed of the Ru $t_{2g}$ orbitals. They are split into the $d_{xz}$ and $d_{yz}$ states that form a nearly double-degenerate spherical Fermi surface sheet [the magenta surface in Fig. 1(c)] and the $d_{xy}$ states producing a cross pattern of three tubes [the blue surface in Fig. 1(c)].

The Fermi surface determines the number of conduction channels, i.e. the number of propagating Bloch states, available for electronic transport. That is determined by

$$N_\parallel^\sigma(\vec{k}_\parallel) = \frac{\hbar}{2}\sum_n \int |v_{nz}^\sigma| \frac{\partial f}{\partial E_n^\sigma(\vec{k})} dk_z, \quad (1)$$

where $\sigma$ denotes the spin index (↑ or ↓), $\vec{k} = (\vec{k}_\parallel, k_z)$ is the wave vector in the Brillouin zone, $\vec{k}_\parallel = (k_x, k_y)$ is the transverse wave vector, $E_n^\sigma$ is energy of band $n$, $v_{nz}^\sigma = \frac{1}{\hbar}\frac{\partial E_n^\sigma}{\partial k_z}$ is the band velocity along the transport $z$ direction, and $f$ is the Fermi distribution function. Figures 2(a) and 2(b) show the calculated number of conduction channels, $N_\parallel^\uparrow$ and $N_\parallel^\downarrow$, for majority and minority spins, respectively, as a function of $\vec{k}_\parallel$ in the 2D Brillouin zone (2DBZ) of SRO (001). The distributions of $N_\parallel^\uparrow$ and $N_\parallel^\downarrow$ reflect the projection of the spin-dependent Fermi surfaces on the plane perpendicular to the transport direction (i.e., [001] in our case), where each Fermi surface sheet adds one conduction channel at a given $\vec{k}_\parallel$ if its projection to this point is non-vanishing [Figs. 2(c,d)]. As a result, for the majority spins, we have $N_\parallel^\uparrow = 1$ in the regions of the 2DBZ where one of the two

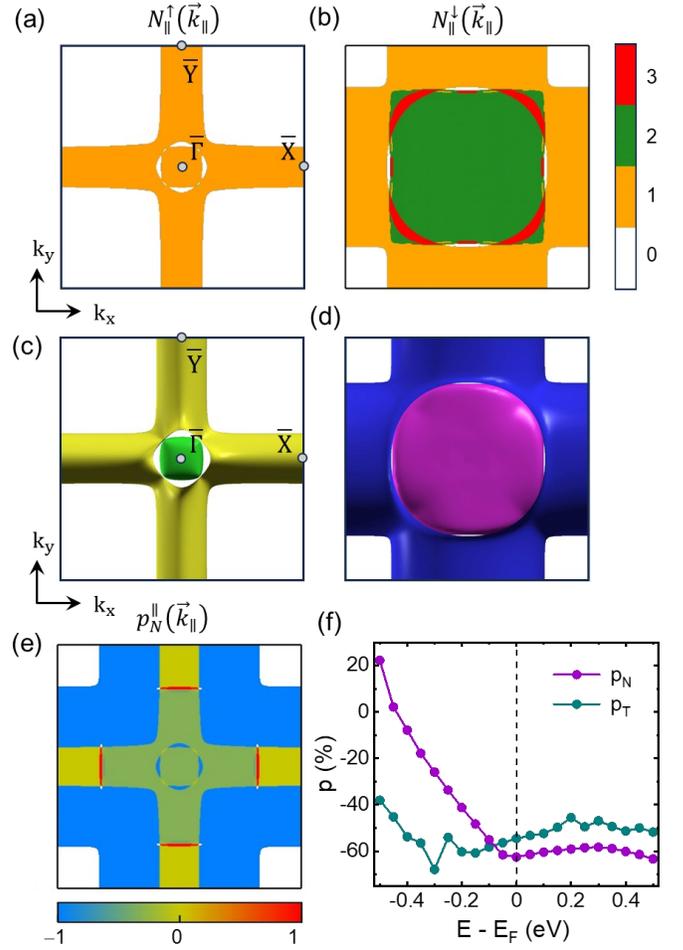

**FIG. 2** (a,b) The number of conduction channels as a function of $\vec{k}_\parallel$ in the 2D Brillouin zone for majority (a) and minority (b) spins at the Fermi energy $E_F$. High-symmetry points are indicated. (c,d) Projection of the majority- (c) and minority- (d) Fermi surface on the (001) plane. (e) Distribution of $\vec{k}_\parallel$-dependent spin polarization $p_N^\parallel(\vec{k}_\parallel)$ in the 2D Brillouin zone at $E_F$. Colors are used to delineate different Fermi surface sheets. (f) The spin polarization of the total number of conduction channels, $p_N$, and the spin polarization of the interface transmission function, $p_T$, as functions of energy. Vertical dashed line indicates the Fermi energy.

non-overlapping Fermi surface sheets, corresponding to the $d_{x^2-y^2}$ and $d_{z^2}$ states, is projected on the (001) plane. For the minority spins, we find $N_{\parallel}^{\downarrow} = 2$ around the $\bar{\Gamma}$ point due to the contribution from the nearly double-degenerate $d_{xz}$ and $d_{yz}$ bands, and $N_{\parallel}^{\downarrow} = 3$ at the edge of these Fermi surface sheets due to their overlap with the $d_{xy}$ bands. The latter are projected on the (001) plane at the periphery of the 2DBZ around the $\bar{X}$ and $\bar{Y}$ points creating regions with $N_{\parallel}^{\downarrow} = 1$.

Due to the spin-dependent Fermi surface of SRO [Figs. 1(b) and 1(c)], the number of conduction channels is spin polarized, i.e. $N_{\parallel}^{\uparrow}$ and $N_{\parallel}^{\downarrow}$ have different values and distribution in the 2DBZ of SRO (001). We therefore define a $\vec{k}_{\parallel}$-dependent spin polarization $p_N^{\parallel}(\vec{k}_{\parallel}) = \frac{N_{\parallel}^{\uparrow} - N_{\parallel}^{\downarrow}}{N_{\parallel}^{\uparrow} + N_{\parallel}^{\downarrow}}$, reflecting the relative difference between $N_{\parallel}^{\uparrow}$ and $N_{\parallel}^{\downarrow}$ at each $\vec{k}_{\parallel}$. As seen from Figure 2(e), there are regions where $N_{\parallel}^{\uparrow} \neq 0$, while $N_{\parallel}^{\downarrow} = 0$, or *vice versa*, resulting in the full spin polarization, $p_N^{\parallel} = \pm 1$ ($\pm 100\%$) (the blue- and red-colored areas). There are also regions around the $\bar{X}$ and $\bar{Y}$ points of the 2DBZ where the spin polarization is zero [the yellow-colored areas in Fig. 2(e)].

Figure 2(f) shows the total spin polarization of the number of conduction channels, $p_N = \frac{N^{\uparrow} - N^{\downarrow}}{N^{\uparrow} + N^{\downarrow}}$, as a function of energy, where $N^{\sigma}$ is the total number of conduction channels, $N^{\sigma} = \frac{1}{(2\pi)^2} \int N_{\parallel}^{\sigma}(\vec{k}_{\parallel}) d\vec{k}_{\parallel}$, We find the spin polarization of about $-62\%$ at the Fermi energy ($E_F$). The negative sign of $p_N$ qualitatively reflects a larger weight of minority spin states at the Fermi energy and agrees with the experimental result [33]. We note, however, that the calculated value of $p_N$ does not take into account the effect of a tunnel barrier that is normally used in the Meservey-Tedrow technique and therefore is not expected to have quantitative agreement with the experiment of Ref. [33].

A more relevant to the calculated $p_N$ is a spin polarization that is measured using point contact Andreev reflection (PCAR) spectroscopy [10,11]. This technique does not require a tunnel barrier and measures the spin polarization of a ferromagnetic metal associated with its ballistic conductance [50]. It should be noted, however, the PCAR technique lacks the ability to determine the sign of spin polarization (in contrast to the Meservey-Tedrow techniques), which is obviously a drawback. The ballistic conductance $G^{\sigma}$ per area and spin is related to the number of conduction channels $N_{\parallel}^{\sigma}(\vec{k}_{\parallel})$ integrated over the transverse momenta:

$$G^{\sigma} = \frac{e^2}{\hbar} \frac{1}{(2\pi)^2} \int N_{\parallel}^{\sigma}(\vec{k}_{\parallel}) d\vec{k}_{\parallel} . \qquad (2)$$

Hence, the calculated $p_N$ provides a proper quantitative measure of the spin polarization of SrRuO$_3$ obtained in an PCAR experiment. According to the available PCAR data, the

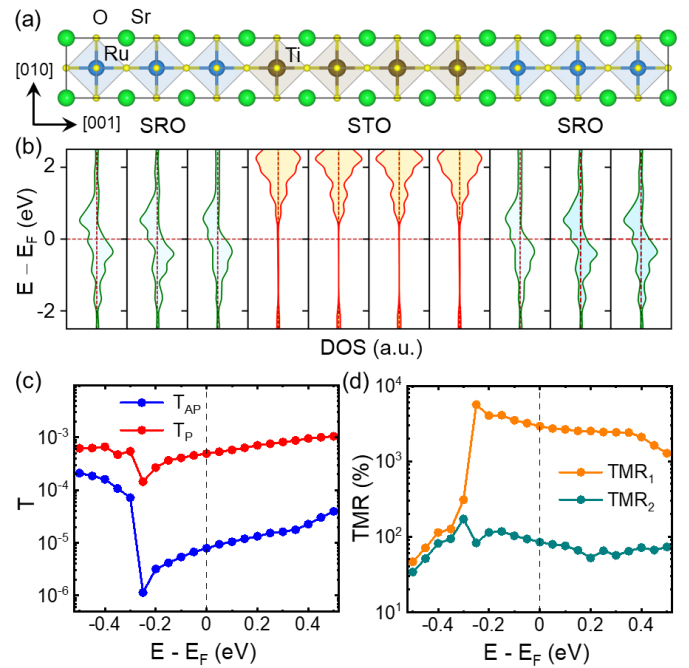

FIG. 3 (a) Atomic structure of SrRuO$_3$/SrTiO$_3$/SrRuO$_3$ MTJ. (b) Layer-dependent density of states (DOS) across the MTJ. The Fermi energy ($E_F$) is indicated by the horizontal dashed line. (c,d) Calculated total transmission of the MTJ for the parallel and antiparallel magnetization states, T$_P$ and T$_{AP}$, respectively, (c) and TMR (d) as a functions of energy. In (d) $TMR_1 = \frac{T_P - T_{AP}}{T_{AP}}$, where T$_P$ and T$_{AP}$ are plotted in (c), whereas $TMR_2 = \frac{2p_T p_T}{1 - p_T p_T}$, where $p_T$ is the spin polarization of the ITF plotted in Fig. 2(f).

measured spin polarization of SrRuO$_3$ ranges from 50 to 60% [51-54], which is in excellent agreement with the calculated value $|p_N| = 62\%$.

As seen from Figure 2(f), the absolute value of $p_N$ is gradually reducing with the decrease of energy and changes sign at around $E = E_F - 0.45$ eV. This behavior is largely follows from the reduced size of the minority-spin Fermi surface area associates with the Ru $t_{2g}$ bands, and the appearance of the majority-spin $t_{2g}$ bands at lower energies [Fig. 1(a)].

## 4. Spin-dependent tunneling in SRO/STO/SRO MTJs

Next, we construct an SRO/STO/SRO (001) MTJ where SRO (001) serves as ferromagnetic electrodes and STO (001) as an insulating tunnel barrier. Figure 3(a) shows the atomic structure of the SRO/STO/SRO (001) heterostructure that is structurally optimized, as described in Sec 2. Our DFT calculations find that a wide band gap of STO is well maintained across the junction and the Fermi energy is located well inside the band gap [Fig. 3(b)], thus providing conditions for direct electron tunneling. This heterostructure serves as a scattering region connected to

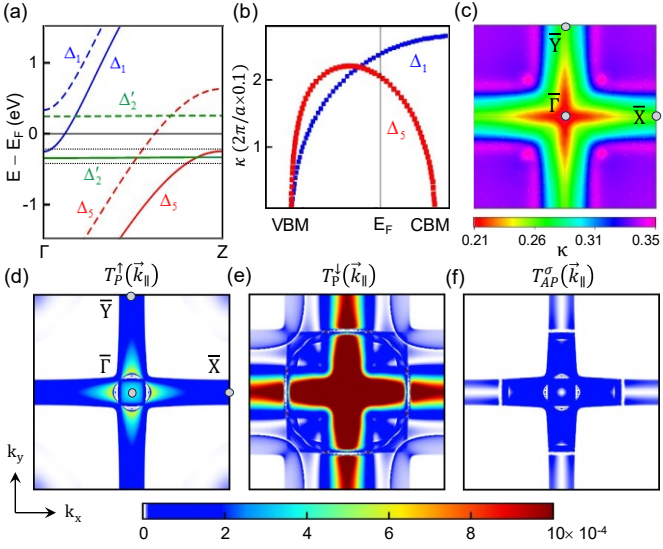

**FIG. 4** (a) Spin-polarized bands along the [001] direction for SrRuO$_3$. Majority-spin (solid) and minority-spin (dashed) bands near the Fermi energy are labeled according to their symmetry. The Fermi energy is set to zero (solid black line). The dotted black lines indicate $E=E_F$ -0.2 eV (the upper line) and $E=E_F$ -0.4 eV (the lower line). (b) Complex bands of SrTiO$_3$ with two lowest decay rates, calculated at the $\bar{\Gamma}$ point. (c) $k_\parallel$-resolved decay rate calculated at Fermi energy of the MTJ. (d-f) $k_\parallel$-resolved transmission at $E = E_F$ for majority- (d) and minority- (e) spin electrons for parallel-aligned MTJ and for either-spin electrons ($\sigma = \uparrow$ or $\downarrow$) for antiparallel-aligned MTJ (f).

two semi-infinite SRO (001) electrodes for our calculations of the transport properties of SRO/STO/SRO (001) MTJ. We calculate spin-resolved transmissions for parallel magnetization ($T_P = T_P^\uparrow + T_P^\downarrow$) and antiparallel magnetization ($T_{AP} = T_{AP}^\uparrow + T_{AP}^\downarrow$, where $T_{AP}^\uparrow = T_{AP}^\downarrow$ by symmetry) of the SRO electrodes in the MTJ. Figure 3(c) shows the results of this calculation as a function of energy $E$. It is seen that for all energies around the Fermi energy and above, $T_P$ is much larger than $T_{AP}$, resulting is a very large TMR ratio $TMR = \frac{T_P - T_{AP}}{T_{AP}}$. At $E = E_F$, we obtain a giant TMR of more than 2900 %. With decreasing the energy, the TMR ratio becomes even larger, reaching the maximum of 5630% at $E = E_F - 0.25$ eV. At lower energies, however, it drops down to about 50% at $E = E_F - 0.5$ eV.

The predicted giant TMR effects in an SRO/STO/SRO (001) MTJ can be explained by considering the symmetry group of wave vector. In crystalline MTJs, where the transverse wave vector $\vec{k}_\parallel$ is conserved during tunneling, the wave functions of the MTJ belong to the symmetry group of the wave vector. This leads to the requirement of symmetry matching between the incoming and outgoing Bloch states in the electrodes and the evanescent states in the barrier. In the following, we therefore analyze the symmetry of the propagating Bloch states in SRO (001) and the evanescent states in STO (001).

In bulk SRO and STO, the cubic crystal field splits the 3$d$-orbitals of the Ru and Ti atoms into higher energy two-fold degenerate $e_g$ bands (formed of the $d_{x^2-y^2}$ and $d_{z^2}$ orbitals) and lower-energy three-fold degenerate $t_{2g}$ bands (formed of $d_{xy}$, $d_{xz}$, and $d_{yz}$ orbitals). In the SRO/STO/SRO (001) MTJ, the symmetry is lowered from cubic to tetragonal thus lifting the partial degeneracy of the $t_{2g}$ and $e_g$ bands: the $t_{2g}$ band splits into a doubly degenerate ($d_{xy}$, $d_{xz}$) band and a non-degenerate $d_{xy}$ band and the $e_g$ band splits into non-degenerate $d_{x^2-y^2}$ and $d_{z^2}$ bands. Along the [001] direction (denoted by $\Delta$) of the layered perovskite structure, the symmetry group of the wave vector is equivalent to that of the $C_{4v}$ point group and has four irreducible representations: $\Delta_1(z^2)$, $\Delta_2(xy)$, $\Delta'_2(x^2-y^2)$, and $\Delta_5(xz, yz)$. The band structure of SRO along the $\Gamma$ - Z symmetry line [Fig. 4(a)] indicates that there are two bands crossing the Fermi energy: the majority-spin band of the $\Delta_1$ symmetry and the doubly degenerate minority-spin band of the $\Delta_5$ symmetry.

For efficient transmission across the STO barrier layer, the symmetry of these propagating Bloch states in SRO (001) needs to be matched to the symmetry of low-decay-rate evanescent states in STO (001). The evanescent states appear within the band gap of STO, characterized by wave-functions that decay exponentially with a rate κ, which is determined by the complex band structure [12-15]. Figure 4 (b) shows the complex bands of STO (001) with the lowest decay rates calculated at the $\bar{\Gamma}$ point ($k_\parallel = 0$) of the 2DBZ. These complex bands represent a $\Delta_5$ doublet and a $\Delta_1$ singlet. The wave-function symmetry must be maintained across the whole crystalline MTJ. As a result, at the $\bar{\Gamma}$ point, the majority-spin states of SRO decay inside the barrier according to the $\Delta_1$ decay rate of STO, whereas the minority-spin states of SRO decay according to the $\Delta_5$ decay rate, giving rise to a perfect correspondence between the band symmetry and spin.

These symmetry constraints lead to a perfect spin-valve effect at the $\bar{\Gamma}$ point. For the parallel-aligned MTJ, the majority-spin states of the $\Delta_1$ symmetry and the minority-spin states of the $\Delta_5$ symmetry are efficiently transmitted from the left to the right SRO electrode across the STO barrier. In contrast, for the antiparallel-aligned MTJ, majority-spin $\Delta_1$ states of the left electrode cannot be transmitted to the minority-spin $\Delta_5$ states of the right electrode and vice versa. Thus, transmission of the antiparallel-aligned MTJ is expected to be zero at the $\bar{\Gamma}$ point.

To explicitly demonstrate these symmetry-driven features, we decompose the total transmission at $E_F$ into the contributions from each transverse wave vector $\vec{k}_\parallel$ and plot $\vec{k}_\parallel$-resolved transmissions $T_P^\uparrow(\vec{k}_\parallel)$, $T_P^\downarrow(\vec{k}_\parallel)$, and $T_{AP}^\sigma(\vec{k}_\parallel)$ ($\sigma = \uparrow$ or $\downarrow$) in Figures 4(d-f), respectively. Note that $T_{AP}^\uparrow(\vec{k}_\parallel) = T_{AP}^\downarrow(\vec{k}_\parallel)$ by

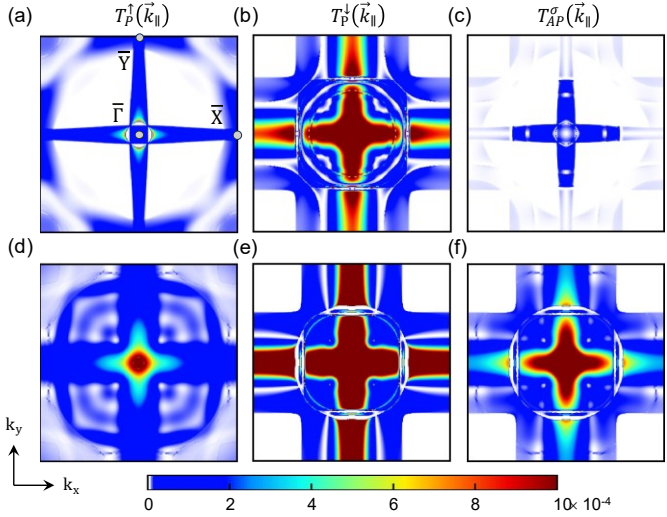

**FIG. 5** (a-f) $k_\parallel$-resolved transmission at $E = E_F - 0.2$ eV (a-c) and $E = E_F - 0.4$ eV (d-f) for majority- (a,d) and minority- (b,e) spin electrons for parallel-aligned MTJ and for either-spin electrons ($\sigma = \uparrow$ or $\downarrow$) electrons for antiparallel-aligned MTJ (c,f).

symmetry. As is evident from the Figure, at the $\bar{\Gamma}$ point ($k_\parallel = 0$), $T_P^\downarrow(\vec{k}_\parallel) \gg T_P^\uparrow(\vec{k}_\parallel)$, and $T_{AP}^\sigma(\vec{k}_\parallel) = 0$. The latter is due to the symmetry mismatch between the $\Delta_1$ majority-spin states of the left electrode and the minority-spin $\Delta_5$ states in the right electrode. The former is due to the much smaller Fermi wave vector of the majority-spin $d_{z^2}$ states compared to that of the $d_{xz}$ and $d_{yz}$ states along the $\Gamma$ - Z symmetry line [Fig. 1 (b,c)].

While this symmetry constraint is not explicitly satisfied away from the $\bar{\Gamma}$ point in the 2DBZ, it is seen from Figures 4(e) and 4(f) that $T_P^\downarrow(\vec{k}_\parallel)$ remains much larger that $T_{AP}^\sigma(\vec{k}_\parallel)$ around the $\bar{\Gamma}$ point, producing a sizable contribution to the overall TMR. It is notable that the transmission distributions in Figures 4(d-f) form cross patterns with the largest contributions along the vertical and horizontal midlines in the 2DBZ. This feature is explained by the calculated $\vec{k}_\parallel$-resolved lowest decay rates of the evanescent states in STO (001). As seen from Figure 4(c), at the Fermi energy, the distribution of the lowest decay rates in the 2DBZ has a pronounced cross pattern resembling that in the transmission distributions in Figures 4 (d-f).

Next, we elucidate the origin of the TMR enhancement at energies below the Fermi energy down to $E = E_F - 0.25$ eV followed by a significant drop in TMR at $E = E_F - 0.3$ eV [Fig. 3(d)]. While the energy dependence of TMR could not be explicitly measured, it is relevant to a voltage dependence of TMR and, more importantly, provides important insights into the physics of spin-dependent tunneling. Figures 5 (a-c) show the calculated $\vec{k}_\parallel$-resolved transmissions at $E = E_F - 0.2$ eV, where we observe significant enhancement of TMR up to about 4000% [Fig. 3(d)]. As seen from by the upper dotted line in Figure 4(a), indicating $E = E_F - 0.2$ eV, there are two bands crossing this energy along the $\Gamma$ - Z line: the majority-spin $\Delta_1$ band and the minority-spin $\Delta_5$ band. These bands are the same as those that appear at the Fermi energy, and hence the symmetry selection rule remains unchanged. However, with reducing the energy closer to the bottom the majority-spin $\Delta_1$ band and to the nearly flat majority-spin $\Delta'_2$ band [Fig. 4(a)], the majority-spin Fermi surface shrinks, reducing the area of available conducting channels for electron tunneling in the antiparallel configuration of MTJ. This substantially reduces $T_{AP}^\sigma(\vec{k}_\parallel)$ and hence enhances TMR. Further reduction of energy down to $E = E_F - 0.4$ eV [the lower dotted line in Figure 4(a)] fully eliminates the majority-spin $\Delta_1$ and $\Delta'_2$ bands from those contributing to transmission. Instead, a majority-spin $\Delta_5$ band appears at this energy and participates in the tunneling process. This band has the same symmetry as the minority-spin $\Delta_5$ band, thus lifting the spin-symmetry mismatch for the antiparallel-aligned MTJ. As a result, and as is evident from Figures 5(d-f), we observe a sizable transmission at the $\bar{\Gamma}$ point and around it for both the $T_P^\uparrow(\vec{k}_\parallel)$ and $T_{AP}^\sigma(\vec{k}_\parallel)$. This leads to a significant reduction of TMR down to about 110% at this energy.

## 5. Spin polarization and its relevance to TMR

Next, we discuss transport spin polarization associated with the spin-dependent tunneling in SRO/STO/SRO (001) MTJs and its relevance to TMR. As we have already mentioned, the spin polarization of the number of conduction channels $p_N$ does not take into account the effect of the tunnel barrier and is more relevant to the spin polarization measured using PCAR spectroscopy [10,11]. The quantity that is normally considered as relevant to TMR is the spin polarization of the interface transmission function (ITF) that includes the ferromagnet/barrier pair rather than the ferromagnet alone [18,19]. Explicitly, the ITF is proportional to the square of the amplitude of the evanescent barrier wavefunction, matched with the scattering state incident from a given electrode, in the middle of the tunnel barrier. If the tunneling current at a given transverse wavevector $\vec{k}_\parallel$ is dominated by a *single* evanescent barrier state, then the $\vec{k}_\parallel$-resolved transmission probability $T^\sigma(\vec{k}_\parallel)$ of the MTJ can be represented as a product of the ITFs $t_i^\sigma(\vec{k}_\parallel)$ for the left ($i = 1$) and right ($i = 2$) electrodes [18, 19]:

$$T^\sigma(\vec{k}_\parallel) = t_1^\sigma(\vec{k}_\parallel) t_2^\sigma(\vec{k}_\parallel). \qquad (3)$$

This approximation becomes better with increasing barrier thickness, because tunneling states with larger decay rates are exponentially suppressed.

In our case, the two electrodes are the same and hence the ITF can be obtained from $T_P^\sigma(\vec{k}_\parallel)$, as follows

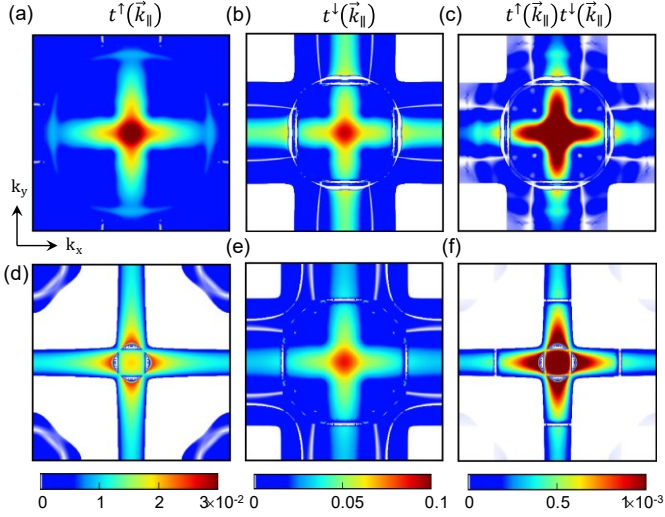

**FIG. 6** (a, b) $k_\parallel$-resolved interface transmission function (ITF) for majority- (a) and minority- (b) spin electrons at $E = E_F - 0.4$ eV. (c) $k_\parallel$-resolved transmission per spin for antiparallel-aligned MTJ calculated from the ITFs at $E = E_F - 0.4$ eV. (d,e) Same as (a,b), respectively, at $E = E_F$. (f) Same as (c) at $E = E_F$.

$$t^\sigma(\vec{k}_\parallel) = \sqrt{T_P^\sigma(\vec{k}_\parallel)}. \quad (4)$$

The fact that transmission can be factorized according to Eq. (3) implies that transmission of the antiparallel-aligned MTJ can be obtained from $t^\sigma(\vec{k}_\parallel)$:

$$T_{AP}^\sigma(\vec{k}_\parallel) = t^\uparrow(\vec{k}_\parallel) t^\downarrow(\vec{k}_\parallel), \quad (5)$$

where $\sigma = \uparrow$ or $\downarrow$. Figures 6(a) and 6(b), respectively, show the $k_\parallel$-resolved majority- and minority-spin ITFs calculated using Eq. (4), and Figure 6(c) shows the $k_\parallel$-resolved $T_{AP}^\sigma(\vec{k}_\parallel)$ calculated using Eq. (6) at $E = E_F - 0.4$ eV. Comparing Figure 6(c) with Figure 5(f), we observe a reasonable agreement between the distributions of the transmission $T_{AP}^\sigma(\vec{k}_\parallel)$ in the 2DBZ calculated explicitly [(Fig. 5(f)] and using the factorization of Eq. (5) [(Fig. 6(c)] at $E = E_F - 0.4$ eV. In contrast, as seen from Figures 6 (d-f), the same calculation performed at $E = E_F$ reveals significant disagreement between the $k_\parallel$-resolved $T_{AP}^\sigma(\vec{k}_\parallel)$ calculated explicitly [(Fig. 4(f)] and using the factorization of Eq. (7) [(Fig. 6(f)] in terms of $t^\sigma(\vec{k}_\parallel)$ [Figs. 6(d,e)]. This difference between the results obtained for different energies ($E = E_F$ and $E = E_F - 0.4$) reflects a change in the transport mechanism as explained next.

As we saw above, at energies $E \leq E_F - 0.3$ eV, the tunneling transmission between SRO (001) electrodes across the STO tunnel barrier is determined by the majority- and minority-spin bands that belong to the $\Delta_5$ symmetry. As a result, *the only* evanescent state in STO (001) that controls transmission (at the $\bar{\Gamma}$ point and around it) also belongs to the $\Delta_5$ symmetry. There is largely no contribution from other evanescent states. This is the condition for the factorization [Eq. (3)] to be valid. Therefore, at energies below $E = E_F - 0.3$ eV, the spin-polarized tunneling in SRO/STO/SRO (001) MTJs can be well described using the concept of ITF. On the contrary, at energies $E \geq E_F - 0.25$ eV, the majority- and minority-spin Bloch states belong to the different symmetries, $\Delta_1$ and $\Delta_5$ respectively. As a result, they are transmitted across the STO barrier through the different evanescent states of the respective symmetries. At these conditions, the factorization (3) fails for the antiparallel-aligned MTJ where two evanescent states of different symmetry are present.

This has important implications for spin polarization as a measure of TMR. As seen from Figures 6(a,b), the ITFs are largely dominated by the cross area around the $\bar{\Gamma}$ point where the decay constant $\kappa(\vec{k}_\parallel)$ has a minimum [Fig. 4(c)]. As a result, at those energies where the $\vec{k}_\parallel$-dependent factorization [Eq. (3)] is valid (i.e., $E \leq E_F - 0.3$ eV), we expect that the factorization of the *total* transmission $T^\sigma$ in terms of the integrated ITF

$$t^\sigma = \frac{1}{(2\pi)^2} \int t^\sigma(\vec{k}_\parallel) d\vec{k}_\parallel, \quad (6)$$

should also be a reasonable approximation. This implies that the transmission for the parallel- and antiparallel-aligned MTJs can be, respectively, written as

$$T_P^\sigma = (t^\sigma)^2, \quad (7)$$

$$T_{AP}^\sigma = t^\uparrow t^\downarrow. \quad (8)$$

This factorization implies that TMR can be described in terms of Julliere's formula applied to MTJ with the same electrodes:

$$TMR = \frac{2 p_T p_T}{1 - p_T p_T}, \quad (9)$$

where $p_T$ is the spin polarization of the total (integrated) interface transmission

$$p_T = \frac{t^\uparrow - t^\downarrow}{t^\uparrow - t^\downarrow}. \quad (10)$$

In Figure 2(f), we plot the calculated spin polarization $p_T$ as a function of energy and the corresponding TMR in Figure 3(d). As seen from Figure 3(d), at energies $E \leq E_F - 0.3$ eV, the explicitly calculated TMR from transmissions $T_P$ and $T_{AP}$ of the whole MTJ [$TMR_1$ in Fig. 3(d)] is in reasonable agreement with the TMR calculated from the spin polarization of the ITF [$TMR_2$ in Fig. 3(d)]. This fact indicates that $p_T$ can serve as a proper measure of TMR for this MTJ at energies $E \leq E_F - 0.3$ eV, where the single evanescent state controls transmission. On the contrary, at energies $E \geq E_F - 0.25$ eV, we observe a huge disagreement between the explicitly calculated TMR and the TMR that is obtained from $p_T$ using Julliere's formula. This is due to the transmission across the STO barrier being controlled by two different evanescent states lifting the condition for the factorization [Eq. (3)], as was discussed above.

Based on this result, we can draw three conclusions. First, $p_T$ can serve as a proper measure of TMR under conditions when only one evanescent state controls tunneling transmission and later is dominated by a region of the 2DBZ where the decay rate is lowest. This is, for example, the case for Fe/MgO/Fe MTJs [13]. However, under conditions where majority- and minority-spin states are transported across the barrier through two different evanescent states, such as at $E = E_F$ in our case, $p_T$ cannot serve as a proper measure of TMR. In principle, one can anticipate a situation where $p_T = 0$, while the TMR is very large driven by the symmetry mismatch between the majority- and minority-spin states. In fact, if the transmission is not factorized, there is no proper quantity, in general, which could be defined as the spin polarization to characterize TMR.

Second, if tunneling occurs from a ferromagnetic metal across an insulator to a non-magnetic metal, which has a featureless spin-degenerate Fermi surface, the integrated ITF [Eq. (6)] of the ferromagnet/insulator pair is expected to control the spin dependence of transmission probability. This kind of junction geometry is similar to the Meservey-Tedrow-type experiment where tunneling occurs to a superconductor [3]. If the ferromagnetic/insulator pair is crystalline SRO/STO, we expect that the spin polarization measured in such experiment should be determined by $p_T$. According to our calculation at $E = E_F$ [Fig. 2(f)], $p_T = -55\%$, which has the same sign but larger magnitude than the spin polarization of $-10\%$ measured in Ref. [33]. We argue that the reason for this disagreement is a lack of high crystallinity of the experimentally fabricated junctions which is critical for obtaining the high degree of spin polarization. It would be helpful to revisit these measurements using high-quality epitaxial junctions which can be grown using modern thin-film deposition techniques.

Third, the factorization of transmission [Eqs. (7-8)] in terms of the integrated ITF [Eq. (6)] factually implies that $\vec{k}_\parallel$ conservation does not any longer hold, and the incoming Bloch states with a given $\vec{k}_\parallel$ can be transmitted to any arbitrary $\vec{k}'_\parallel$. This behavior is relevant to MTJs where non-resonant localized states in the barrier lead to diffuse scattering. For example, this is the case for Fe/MgO/Fe MTJs which contain O vacancies [55]. We argue that diffuse scattering by O vacancies may be relevant to the recent experiments of Ref. [34]. It is well known that the formation energy of O vacancies in STO is low, and thus they are likely present in the STO tunnel barrier (if not specially controlled). Under conditions of diffuse scattering the predicted TMR is 85% [Fig. 3(d) at $E = E_F$], which is somewhat larger but comparable to that measured experimentally (25%).

## 6. Conclusions

Using first-principles quantum-transport calculations, we have investigated spin-polarized transport properties of crystalline $SrRuO_3/SrTiO_3/SrRuO_3$ (001) MTJs and predicted a giant TMR effect of nearly 3000%. This giant TMR is driven by symmetry matching (mismatching) of the incoming and outcoming Bloch states in the SRO (001) electrodes and evanescent states in the STO (001) barrier. We argued that under the conditions of symmetry-controlled transport in these MTJs, spin polarization, whatever definition is used, is not a relevant measure of spin-dependent tunneling. In the presence of diffuse scattering, however, e.g. due to localized states in the band gap of the tunnel barrier produced by O vacancies, symmetry matching is no longer valid and TMR in SRO/STO/SRO (001) MTJs is strongly reduced. Under these conditions, the spin polarization of the SRO/STO (001) interface transmission function becomes a valid measure of TMR. These results provide an important insight into understanding and optimizing TMR in all-oxide MTJs. In particular, it is likely that symmetry-controlled tunneling predicted in this work has not yet been practically realized in experiments due O oxygen vacancies and/or other defects present in the tunnel barrier. We therefore encourage experimentalists working in this field to improve quality of their SRO/STO/SRO (001) MTJs to search for the giant TMR effects predicted in this work.

## Acknowledgments

This work was primarily supported by the Division of Materials Research of the National Science Foundation (NSF grant No. DMR-2316665). The authors also acknowledge support from a UNL Grand Challenges catalyst award entitled Quantum Approaches Addressing Global Threats. Computations were performed at the University of Nebraska-Lincoln Holand Computing Center and Hefei Advanced Computing Center.

* ksamanta2@unl.edu
** tsymbal@unl.edu